\documentclass[epsfig]{revtex4}
\usepackage{graphicx}
\begin{document}

\title{Genuine ($k$, $m$)-threshold controlled teleportation and its security}
\author{Xin-Wen Wang$^{1}$, Da-Chuang Li$^{2}$, and Guo-Jian Yang$^{1}$\footnote{E-mail: yanggj@bnu.edu.cn}}
\address{$^1$Department of Physics, and Applied Optics Beijing Area Major Laboratory, Beijing Normal University, Beijing
100875, China\\
$^2$School of Physics and Material Science, Anhui University, Hefei
230039, China}

\begin{abstract}
We propose genuine ($k$, $m$)-threshold controlling schemes for
controlled teleportation via multi-particle entangled states, where
the teleportation of a quantum state from a sender (Alice) to a
receiver (Bob) is under the control of $m$ supervisors such that $k$
($k\leq m$) or more of these supervisors can help Bob recover the
transferred state. By construction, anyone of our quantum channels
is a genuine multipartite entangled state of which any two parts are
inseparable. Their properties are compared and contrasted with those
of the well-known Greenberger-Horne-Zeilinger, W, and linear cluster
states, and also several other genuine multipartite entangled states
recently introduced in literature. We show that our schemes are
secure against both Bob's dishonesty and supervisors' treacheries.
For the latter case, the game theory is utilized to prove that
supervisors' cheats can be well prevented. In addition to their
practical importance, our schemes are also useful in seeking and
exploring genuine multipartite entangled states and opening another
perspective for the applications of the game theory in quantum
information science.

 PACS number(s): 03.67.Dd, 03.67.Hk, 03.67.Mn

\end{abstract}

\maketitle

\section{introduction}
In quantum information science, information is encoded in quantum
states. Quantum information processing is in fact the manipulation
or (and) transfer of quantum states. Quantum teleportation
\cite{70PRL1895} is a typical quantum information processing task,
which functions as transferring a quantum state from one site to
another one via previously shared entanglement assisted by classical
communications and local operations. Quantum teleportation can not
only be directly used to realize quantum communication but also
construct a primitive of a quantum computer \cite{primitive}.
Quantum teleportation has been realized in many experiments (see
e.g., \cite{390N575}). Since the end of last century, a new quantum
teleportation idea, i.e., controlled teleportation (CT), has been
attracting much interest
\cite{58PRA4394,70PRA022329,72PRA022338,75PRA052306,40JPB1767,0609026,68PRA022321,341PLA55}.
CT functions as teleporting a quantum state from a sender's (Alice)
site to a receiver's (Bob) site under the control of multiple
supervisors (Charlie 1, Charlie 2,$\cdots$). In other words, Alice
and Bob need the cooperation of Charlies in order to realize the
teleportation (communication) successfully. A CT scheme has already
been demonstrated in an optical experiment \cite{430N54}.

CT is useful in the context of networked quantum communication,
quantum computation, and cryptographic conference
\cite{72PRA022338,59PRA1829,79PRA062315,79PRA062313,08011544,0401076,cryptography}.
For instance, CT can be used as a secret sharing to hide a quantum
state as a secret \cite{72PRA022338,79PRA062313}. In addition, CT
has many similarities with the secure multi-party quantum
computation (MPQC) protocol \cite{08011544} which allows multiple
players to compute an agreed quantum circuit where each player has
access only to his own quantum input. A MPQC protocol has two
phases, sharing phase and reconstruction phase. In the sharing
phase, dealers provide many agents with their initial state; in the
reconstruction phase, one agent is designated to reconstruct the
final state of the protocol with the help of the other ones. CT may
have other interesting applications, such as in opening a credit
account on the agreement of multiple managers in a quantum network.

The previous CT schemes
\cite{58PRA4394,70PRA022329,72PRA022338,75PRA052306,40JPB1767,0609026,68PRA022321,341PLA55}
are focused on the ($m$, $m$)-threshold controlling schemes where
the achievement of teleportation is conditioned on the collaboration
of all the supervisors. In other words, it is impossible to realize
teleportation between Alice and Bob if anyone of Charlies does not
cooperate for subjective or objective reasons. However, a more
general CT scheme should consider the ($k$, $m$)-threshold case
($k\leq m$) where $k$ or more of the supervisors can help Bob
successfully recover the transferred state, but less than $k$ of
them cannot. Recently, different ($k$, $m$)-threshold controlling
schemes were discussed in Refs.~\cite{Wang,79PRA062313}. The scheme
in Ref.~\cite{Wang} needs lowering the fidelity of teleportation and
its successful probability for enduring the uncooperation of part of
supervisors. In Ref.~\cite{79PRA062313}, authors pointed out that a
($k$, $m$)-threshold controlling scheme can be constructed by using
secret sharing. That is, the teleportation is controlled by a
classical key which is shared by the supervisors such that $k$ or
more of them can recover the key. However, as mentioned in
Ref.~\cite{79PRA062313}, a classical key can be easily copied, and
Charlies cannot stop Bob from recovering Alice's original state if
Bob manages to obtain as least $k$ shares of the key without consent
of Charlies. More importantly, the classical ($k$, $m$)-threshold
controlling scheme can not prevent Charlies' cheats as will be
shown. They also proposed another ``($k$, $m$)-threshold'' CT scheme
which is a combination of a ($m$, $m$)-threshold CT scheme and the
($k$, $m$)-threshold secret sharing scheme mentioned above.
Evidently, it is not a genuine ($k$, $m$)-threshold controlling
scheme, because Bob still needs the assistance of all the
supervisors for recovering Alice's original state. In principle, a
($k$, $m$)-threshold controlling scheme can be constructed by using
the quantum polynomial codes \cite{83PRL648} as mentioned in
Ref.~\cite{79PRA062313}. However, it needs the supervisors and Bob
to come together and perform nonlocal operations (multi-particle
joint operations).

In this article, we propose genuine ($k$, $m$)-threshold controlling
schemes for CT. In these schemes, the supervisors (Charlies) only
need to perform single-particle measurements and announce their
outcomes. If the recipient receives $k$ correct outcomes, he or she
can reconstruct the original state that the sender wants to transfer
by appropriate local operations. We first consider the CT of a
single-particle state via a multipartite entangled sate. Then the CT
of an $n$-particle state can be directly realized by using $n$ such
multipartite entangled states. However, the directly generalized
method requires considerably large auxiliary particle resources and
local operations, as well as classical communications, especially
when $n$ is very large. We propose a much more economical scheme for
CT of an arbitrary $n$-particle state with a single multipartite
entangled state. By construction, our quantum channels are genuine
multipartite entangled states in which any two parts are
inseparable. Their properties are compared and contrasted with those
of the well-known Greenberger-Horne-Zeilinger, W, and linear cluster
states, and also several other genuine multipartite entangled states
recently introduced in literature. We show that our schemes are
secure against both Bob's dishonesty and supervisors' treacheries.
For the latter case, the game theory is utilized to prove that
supervisors' cheats can be well prevented. In addition to the
potential applications in networked quantum communication and
quantum computation, our schemes are also useful in seeking and
exploring genuine multipartite entangled states and opening another
perspective for the applications of the game theory in quantum
information science.

The paper is organized as follows. In Sec. II, we describe the ($k$,
$m$)-threshold CT protocols, and briefly analyze the features of the
entanglement channels. In Sec. III, we discuss the security of our
schemes against Bob's dishonesty and supervisors' treacheries.
Concluding remarks appear in Sec. IV.

\section{($k$, $m$)-threshold controlling scheme for controlled teleportation}

\subsection{A brief review of the teleportation scheme with a Bell
state} Quantum teleportation was first proposed by Bennett \emph{et
al.} \cite{70PRL1895}. In their original scheme, the state to be
teleported is an arbitrary single-particle state given by
\begin{equation}
\label{psi}
 |\psi\rangle_T=\alpha|0\rangle_T+\beta|1\rangle_T
\end{equation}
with $|\alpha|^2+|\beta|^2=1$, and the quantum channel shared by the
sender Alice and the receiver Bob is an EPR singlet state. In fact,
the quantum channel can be anyone of the four Bell basis states
\begin{eqnarray}
\label{Bell}
 |\mathcal{B}^1\rangle_{AB}=\frac{1}{\sqrt{2}}(|00\rangle+|11\rangle)_{AB},\nonumber\\
 |\mathcal{B}^2\rangle_{AB}=\frac{1}{\sqrt{2}}(|00\rangle-|11\rangle)_{AB},\nonumber\\
 |\mathcal{B}^3\rangle_{AB}=\frac{1}{\sqrt{2}}(|01\rangle+|10\rangle)_{AB},\nonumber\\
 |\mathcal{B}^4\rangle_{AB}=\frac{1}{\sqrt{2}}(|01\rangle-|10\rangle)_{AB}.
\end{eqnarray}
Note that the four Bell states can be transformed into each other by
local operations on one particle. For instance,
$|\mathcal{B}^1\rangle_{AB}=\sigma^z_B|\mathcal{B}^2\rangle_{AB}
 =\sigma^x_B|\mathcal{B}^3\rangle_{AB}=i\sigma^y_B|\mathcal{B}^4\rangle_{AB}$,
 where $\sigma^j$ ($j=x,y,z$) are the conventional Pauli matrices given by
\begin{eqnarray}
 \sigma^x=\left(
 \begin{array}{cc}
  0 & 1 \\ 1 & 0 \\
\end{array}
\right),~
 \sigma^y=\left(
 \begin{array}{cc}
  0 & -i \\ i & 0 \\
\end{array}
\right),~
 \sigma^z=\left(
 \begin{array}{cc}
  1 & 0 \\ 0 & -1 \\
\end{array}
\right).
\end{eqnarray}
As an example, we assume that the quantum channel is
$|\mathcal{B}^1\rangle_{AB}$. Then the state of the total system is
\begin{eqnarray}
\label{Psitotal}
 |\Psi\rangle_{total}&=&|\psi\rangle_T\otimes|\mathcal{B}^1\rangle_{AB} \nonumber\\
 &=&\frac{1}{2}\left[|\mathcal{B}^1\rangle_{TA}(\alpha|0\rangle +\beta|1\rangle)_B \right.\nonumber\\
 &&+ |\mathcal{B}^2\rangle_{TA}(\alpha|0\rangle -\beta|1\rangle)_B\nonumber\\
 &&+|\mathcal{B}^3\rangle_{TA}(\alpha|1\rangle + \beta|0\rangle)_B\nonumber\\
 &&\left.+ |\mathcal{B}^4\rangle_{TA}(\alpha|1\rangle - |0\rangle)_B\right]\nonumber\\
&=&\frac{1}{2}\left[|\mathcal{B}^1\rangle_{TA}|\psi\rangle_B
              +\sigma_A^z|\mathcal{B}^1\rangle_{TA}\sigma_B^z|\psi\rangle_B\right.\nonumber\\
 &&\left.+\sigma_A^x|\mathcal{B}^1\rangle_{TA}\sigma_B^x|\psi\rangle_B
   +(-i\sigma_A^y)|\mathcal{B}^1\rangle_{TA}(-i\sigma_B^y)|\psi\rangle_B\right].
\end{eqnarray}
Alice performs a Bell-basis measurement on particles $T$ and $A$ and
broadcasts the outcomes, after which Bob applies the required Pauli
rotation to transform the state of his particle $B$ into an accurate
replica of the original state of Alice's particle $T$. The
one-to-one correspondence between Alice's possible measurement
outcomes and the required Pauli rotations can be easily obtained
from Eq.~(\ref{Psitotal}). It can be easily proved that if the
quantum channel is another Bell state $|\mathcal{B}^j\rangle_{AB}$
($j=2$, 3, or 4), the state of the total system can also be expanded
as
\begin{eqnarray}
|\Psi\rangle_{total}&=&|\psi\rangle_T\otimes|\mathcal{B}^j\rangle_{AB} \nonumber\\
        &=&\frac{1}{2}\left[|\mathcal{B}^j\rangle_{TA}|\psi\rangle_B
       +\sigma_A^z|\mathcal{B}^j\rangle_{TA}\sigma_B^z|\psi\rangle_B\right.\nonumber\\
 && \left.+\sigma_A^x|\mathcal{B}^j\rangle_{TA}\sigma_B^x|\psi\rangle_B
     +(-i\sigma_A^y)|\mathcal{B}^j\rangle_{TA}(-i\sigma_B^y)|\psi\rangle_B\right].
\end{eqnarray}
Thus the one-to-one correspondence between Alice's possible
measurement outcomes and the required Pauli rotations can always be
easily obtained.

\subsection{($k$, $m$)-threshold controlled teleportation for an arbitrary single-particle state}
Before discussing the ($k$, $m$)-threshold schemes, we first give a
general description on the basic idea of CT. Assume that there is a
community which is composed of $m+2$ members, Alice, Bob, Charlie 1,
Charlie 2, $\cdots$, and Charlie $m$. The members are distributed in
a network and connected by a quantum channel, i.e., a multipartite
entangled state, and one or more classical channels (can be
considered as the conventional classical communication facilities).
One of Alice and Bob is the sender of a quantum state (the carrier
of quantum information), and the other one is the receiver. Charlies
act as the supervisors who can decide whether or not to allow Alice
and Bob to carry out the teleportation. In a word, the teleportation
of a quantum state between Alice and Bob is supervised by Charlies
and needs their approval. Without loss of generality, we assume
Alice is the sender and Bob is the receiver. In order to realize the
CT of the single-particle state $|\psi\rangle_T$, the quantum
channel shared by them can be in the form of
\begin{eqnarray}
\label{Phi}
 |\Phi\rangle_{2+m}&=&x_1|\mathcal{B}^1\rangle_{AB}|\phi^1\rangle_{C_1C_2\cdots C_m}
    + x_2|\mathcal{B}^2\rangle_{AB}|\phi^2\rangle_{C_1C_2\cdots C_m}\nonumber\\
  && +x_3|\mathcal{B}^3\rangle_{AB}|\phi^3\rangle_{C_1C_2\cdots C_m}
   +x_4|\mathcal{B}^4\rangle_{AB}|\phi^4\rangle_{C_1C_2\cdots C_m},
\end{eqnarray}
where $\sum_{i=1}^4|x_i|^2=1$, $|\phi^i\rangle_{C_1C_2\cdots C_m}$
are normalized and their forms depend on the concrete schemes but
should satisfy $\langle\phi^{i'}|\phi^i\rangle=\delta_{ii'}$ and can
be distinguished by local measurements and classical communications.
Here, particle $A$ belongs to Alice, particle $B$ to Bob, and
particle $C_j$ to Charlie $j$ ($j=1,2,\cdots,m$). It has been shown
in the above subsection that anyone of the four Bell states can be
competent for realizing the teleportation of the state
$|\psi\rangle_T$. However, Alice and Bob can carry out the
teleportation only if they can ascertain which Bell state their
subsystem is in. With the quantum channel $|\Phi\rangle_{2+m}$, the
identification of the Bell states can be achieved by the following
method: Charlies make measurements with appropriate bases on their
own particles and inform Bob the outcomes; then Bob can distinguish
the states $\{|\phi^i\rangle_{C_1C_2\cdots C_m}, i=1,2,3,4\}$ and
thus can identify the Bell states. The one-to-one correspondence
between $\{|B^i\rangle_{AB}\}$ and $\{|\phi^i\rangle_{C_1C_2\cdots
C_m}\}$ is clearly shown in Eq.~(\ref{Phi}). Without the cooperation
of Charlies, the subsystem of Alice and Bob will be in the mixed
state $\rho_{AB}=\mathrm{tr}_{C_1C_2\cdots
C_m}\left(|\Phi\rangle_{2+m}\langle\Phi|\right)
  =|x_1|^2|\mathcal{B}^1\rangle_{AB}\langle\mathcal{B}^1|+|x_2|^2|\mathcal{B}^2\rangle_{AB}\langle\mathcal{B}^2|
  +|x_3|^2|\mathcal{B}^3\rangle_{AB}\langle\mathcal{B}^3|+|x_4|^2|\mathcal{B}^4\rangle_{AB}\langle\mathcal{B}^4|$.
The mixed state cannot be used to implement perfect teleportation
\cite{60PRA1888}.

In the conventional CT schemes which use the
Greenberger-Horne-Zeilinger (GHZ)-type entangled states \cite{GHZ}
as the quantum channel, two terms of $\{x_i,i=1,2,3,4\}$ are set to
zero, and the other two are not and their corresponding
$|\phi\rangle_{C_1C_2\cdots C_m}$ states are different Dicke states.
For example, the quantum channel $|\Phi\rangle_{2+m}$ is a GHZ state
$|GHZ\rangle_{2+m}=(1/\sqrt{2})(|0000\cdots 0\rangle+|1111\cdots
1\rangle)_{ABC_1C_2\cdots C_m}$, then $x_3=x_4=0$,
$x_1=x_2=1/\sqrt{2}$, $|\phi^1\rangle_{C_1C_2\cdots
C_m}=(1/\sqrt{2^m})\left[\sum_{l=0}^{m^+}S_m^{2l}|-\rangle^{\otimes
2l}|+\rangle^{\otimes(m-2l)}\right]$, and
$|\phi^2\rangle_{C_1C_2\cdots
C_m}=(1/\sqrt{2^m})\left[\sum_{l=0}^{m^-}S_m^{2l+1}|-\rangle^{\otimes
  (2l+1)}|+\rangle^{\otimes(m-2l-1)}\right]$, where $|\pm\rangle=(|0\rangle\pm|1\rangle)/\sqrt{2}$,
$S_m^{\tilde{l}}=m!/[\tilde{l}!(m-\tilde{l})!]$
($\tilde{l}=2l,2l+1$) is the combinational coefficient,
$|-\rangle^{\otimes \tilde{l}}|+\rangle^{\otimes(m-\tilde{l})}$
denotes that $\tilde{l}$ particles are in the state $|-\rangle$ and
$m-\tilde{l}$ particles are in the state $|+\rangle$, and when $m$
is odd $m^-=m^+=(m-1)/2$, otherwise, $m^-=m/2-1$ and $m^+=m/2$. That
is, $|\phi^1\rangle_{C_1C_2\cdots C_m}$ and
$|\phi^2\rangle_{C_1C_2\cdots C_m}$ are the Dicke states with even
$|-\rangle$ and odd $|-\rangle$, respectively. Thus Charlies can
perform single-particle measurements on their own particles with the
basis $\{|\pm\rangle\}$ and inform Bob the outcomes, and Bob can
identify the Bell states with the outcomes, even or odd $|-\rangle$.
Evidently, such a CT scheme is a ($m$, $m$)-threshold controlling
scheme, i.e., Alice and Bob can implement the teleportation if and
only if all Charlies agree and cooperate.

Now, let us move on to the ($k$, $m$)-threshold controlling scheme.
For simplicity, we first consider the case $k=1$. That is, Alice and
Bob can realize successfully teleportation if anyone of Charlies
cooperate with them. We can set $x_3=x_4=0$, $x_1=x_2=1/\sqrt{2}$,
$|\phi^1\rangle_{C_1C_2\cdots C_m}=|00\cdots 0\rangle_{C_1C_2\cdots
C_m}$, and $|\phi^2\rangle_{C_1C_2\cdots C_m}=|11\cdots
1\rangle_{C_1C_2\cdots C_m}$ in Eq.~(\ref{Phi}). Then the quantum
channel is
\begin{eqnarray}
\label{Phi1}
 |\Phi^1\rangle_{2+m}&=&\frac{1}{\sqrt{2}}\left(|\mathcal{B}^1\rangle_{AB}|00\cdots 0\rangle_{C_1C_2\cdots C_m}
  +|\mathcal{B}^2\rangle_{AB}|11\cdots 1\rangle_{C_1C_2\cdots C_m}\right)\nonumber\\
 &=& \frac{1}{2}(|0000\cdots 0\rangle+|0011\cdots 1\rangle+|1100\cdots 0\rangle
 -|1111\cdots 1\rangle)_{ABC_1C_2\cdots C_m}.
\end{eqnarray}
It can be seen that if anyone of Charlies performs a measurement on
his particle with the basis $\{|0\rangle,|1\rangle\}$ (i.e., in the
$z$ direction) and informs Bob the outcome, Bob can know particles
$A$ and $B$ are in the Bell state $|\mathcal{B}^1\rangle_{AB}$ for
the outcome $|0\rangle$ or $|\mathcal{B}^2\rangle_{AB}$ for
$|1\rangle$. In other words, anyone of Charlies suffices to help
Alice and Bob achieve the teleportation of the state
$|\psi\rangle_T$. However, if all of Charlies do not collaborate
with them, they cannot achieve the teleportation. Note that any
combination of $\{|00\cdots 0\rangle_{C_1C_2\cdots C_m},|11\cdots
1\rangle_{C_1C_2\cdots C_m}\}$ with two of the four Bell states can
construct a quantum channel which can realize the (1, $m$)-threshold
CT mentioned above. For instance, we can also construct a suitable
quantum channel by setting $x_1=x_2=0$,
$|\phi^3\rangle_{C_1C_2\cdots C_m}=|00\cdots 0\rangle_{C_1C_2\cdots
C_m}$, and $|\phi^4\rangle_{C_1C_2\cdots C_m}=|11\cdots
1\rangle_{C_1C_2\cdots C_m}$ in Eq.~(\ref{Phi}).

For the case $k>1$, the quantum channel can be constructed as
\begin{eqnarray}
\label{Phik}
|\Phi^k\rangle_{2+m}&=&\frac{1}{\sqrt{2}}\left(|\mathcal{B}^1\rangle_{AB}|\phi^1\rangle_{C_1C_2\cdots
   C_m} + |\mathcal{B}^2\rangle_{AB}|\phi^2\rangle_{C_1C_2\cdots C_m}\right)\nonumber\\
|\phi^1\rangle_{C_1C_2\cdots C_m}&=&|00\cdots 0\rangle_{C_1C_2\cdots C_m}\nonumber\\
|\phi^2\rangle_{C_1C_2\cdots
 C_m}&=&\frac{1}{\sqrt{S_m^{k-1}}}|k-1,m-k+1\rangle_{C_1C_2\cdots C_m},
\end{eqnarray}
where $S_m^{k-1}=m!/[(m-k+1)!(k-1)!]$ is the combinational
coefficient, $|k-1,m-k+1\rangle_{C_1C_2\cdots C_m}$ denotes all the
totally symmetric states including $k-1$ zeros and $m-k+1$ ones. For
example, $m=3$ and $k=2$, then $|\phi^2\rangle_{C_1C_2
 C_3}=(1/\sqrt{3})(|011\rangle+|101\rangle+110\rangle)_{C_1C_2
 C_3}$. As a matter of fact, $|\phi^2\rangle_{C_1C_2\cdots
 C_m}$ is then a symmetric Dicke state with $m-k+1$ excitations. By the way,
the symmetric six-qubit Dicke state with three excitations has
recently been realized in experiment \cite{103PRL020503}. Note that
when $k=1$, the state of Eq.~(\ref{Phik}) reduces to that of
Eq.~(\ref{Phi1}). We consider that $l$ ($l\leq m$) of Charlies
perform single-particle measurements on their own particles with the
basis $\{|0\rangle, |1\rangle\}$. There are two cases. (a) $l\geq
k$, if all of them get the outcome $|0\rangle$, the subsystem of
particles $A$ and $B$ collapses into $|\mathcal{B}^1\rangle_{AB}$,
otherwise, it collapses into $|\mathcal{B}^2\rangle_{AB}$. (b)
$l<k$, if all of them get the outcome $|0\rangle$, the subsystem of
particles $A$ and $B$ collapses into a mixed state of
$|\mathcal{B}^1\rangle_{AB}$ and $|\mathcal{B}^2\rangle_{AB}$. Thus
we can conclude that $k$ or more of Charlies can help Alice and Bob
deterministically distinguish between the two Bell states
$|\mathcal{B}^1\rangle_{AB}$ and $|\mathcal{B}^2\rangle_{AB}$, while
less than $k$ of them cannot. In other words, Alice can
deterministically teleport the state $\psi\rangle_T$ to Bob if and
only if $k$ or more of Charlies collaborate with them. The procedure
of such a CT protocol is as follows.

(i) Alice performs a Bell-basis measurement on particles $T$ and
$A$, and informs Bob the outcome, one of
$\{|\mathcal{B}^1\rangle_{TA},|\mathcal{B}^2\rangle_{TA},|\mathcal{B}^3\rangle_{TA},|\mathcal{B}^4\rangle_{TA}\}$.

(ii) Bob sends his petition to Charlies.

(iii)Charlies talk over whether or not to allow Bob to recover the
original state of Alice's particle $T$. If more than a certain
number of Charlies (e.g., $2/3$ of them) vote for allowing, a
collective decision should be made that permitting Bob to recover
Alice's original state. Then all Charlies should perform
single-particle measurements on their own particles with the basis
$\{|0\rangle,|1\rangle\}$ and broadcast their outcomes.

(iv) According to Alice's and Charlies' measurement outcomes, Bob
performs a corresponding Pauli rotation on particle $B$ and recovers
Alice's original state on it.

Note that we need all of Charlies instead of $k$ of them to
broadcast their outcomes in step (iii) is based on the consideration
that there may exist treacherous Charlies who will cheat Bob and
send him the false measurement outcomes. The detailed proof for the
security of our scheme against Charlies' cheats will be given in
Sec. III.

\subsection{($k$, $m$)-threshold controlled teleportation for an arbitrary multi-particle state}
As a direct generalization of the teleportation of a single-particle
state, teleportation of an arbitrary $n$-particle state
\begin{equation}
\label{psiTn}
 |\psi\rangle_{T_1T_2\cdots T_n}=\sum\limits_{j_1,j_2,\cdots,j_n=0}^{1}y_{j_1j_2\cdots j_n}|j_jj_2\cdots
j_n\rangle_{T_1T_2\cdots T_n}
\end{equation}
can be achieved with $n$ Bell states. In fact, the teleportation of
a two-particle state with two Bell states has already been
demonstrated in an optical experiment \cite{0609129}.

Thus, one can use $n$ copies of the state of Eq.~(\ref{Phi}) to
realize the CT of an arbitrary $n$-particle state. Also, we can
directly use $n$ copies of the state $|\Phi^1\rangle_{2+m}$ or
$|\Phi^k\rangle_{2+m}$ [see Eqs.~(\ref{Phi1}) and (\ref{Phik})] to
accomplish the ($k$, $m$)-threshold CT of
$|\psi\rangle_{T_1T_2\cdots T_n}$. However, this method requires
considerably large auxiliary particle resources and local
operations, as well as classical communications, especially when the
number of ``teleported'' qubits is very large. Particularly, each
Charlie needs to hold $n$ controlling particles, perform $n$
single-particle measurements, and send Bob $n$ bits of classical
information about the measurement outcomes.

We now propose a much more economical way to implement the ($k$,
$m$)-threshold CT of an arbitrary $n$-particle state. The quantum
channel is the multipartite entangled state
\begin{eqnarray}
\label{Phik2nm}
 |\Phi^k\rangle_{2n+m}&=&\frac{1}{\sqrt{2}}\left(\prod\limits_{i=1}^{n}|\mathcal{B}^1\rangle_{A_iB_i}
      \otimes|00\cdots 0\rangle_{C_1C_2\cdots C_m}\right.\nonumber\\
&&\left.+\prod\limits_{i=1}^{n}|\mathcal{B}^2\rangle_{A_iB_i}\otimes\frac{1}{\sqrt{S_m^{k-1}}}|k-1,m-k+1\rangle_{C_1C_2\cdots C_m}\right),\nonumber\\
\end{eqnarray}
where particles $A_i$ are held by Alice, $B_i$ held by Bob. In order
to successfully implement the teleportation, Alice and Bob need
Charlies to help them identify the two sequences of Bell states.
Particularly, the procedure is as follows.

(i) Alice performs a sequence of Bell-basis measurements on the pairs
of particles $\{(T_i,A_i),i=1,2,\cdots,n\}$, and informs Bob the
outcomes.

(ii) and (iii) are the same as that of the CT protocol for a
single-particle state.

(iii) According to Alice's and Charlies' measurement outcomes, Bob
applies the corresponding Pauli rotations on particles
$\{B_i,i=1,2,\cdots,n\}$ and reconstructs the state of
Eq.~(\ref{psiTn}).

 As shown above, regardless of the number of
qubits to be teleported, the proposed approach only requires that
each supervisor holds one particle, performs one single-particle
measurement on his or her particle, and send one bit of classical
message to the receiver Bob. Therefore, compared with the directly
generalized method mentioned above, this method is much more
economical, because the required auxiliary particle resources, the
number of measurements, and the quantity of classical communications
are greatly reduced.

We notice that any two of the four Bell states can be distinguished
by local (single-particle) measurements with appropriate measurement
bases and classical communications. For instance, we can distinguish
between the two sets
$\{|\mathcal{B}^1\rangle,|\mathcal{B}^2\rangle\}$ and
$\{|\mathcal{B}^3\rangle,|\mathcal{B}^4\rangle\}$ by using the
measurement basis $\{|0\rangle,|1\rangle\}$, which can be evidently
seen from Eq.~(\ref{Bell}). In order to show how to distinguish
between the two sets
$\{|\mathcal{B}^1\rangle,|\mathcal{B}^3\rangle\}$ and
$\{|\mathcal{B}^2\rangle,|\mathcal{B}^4\rangle\}$ by local
measurements and classical communication, we rewrite them as
\begin{eqnarray}
\label{Bell1}
 |\mathcal{B}^1\rangle=\frac{1}{\sqrt{2}}(|++\rangle+|--\rangle),\nonumber\\
 |\mathcal{B}^2\rangle=\frac{1}{\sqrt{2}}(|-+\rangle+|+-\rangle),\nonumber\\
 |\mathcal{B}^3\rangle=\frac{1}{\sqrt{2}}(|++\rangle-|--\rangle),\nonumber\\
 |\mathcal{B}^4\rangle=\frac{1}{\sqrt{2}}(|-+\rangle-|+-\rangle).
\end{eqnarray}
Obviously, if two participants perform, respectively, a
single-particle measurement on different particles with the basis
$\{|\pm\rangle\}$, they can discriminate between the two sets
$\{|\mathcal{B}^1\rangle,|\mathcal{B}^3\rangle\}$ and
$\{|\mathcal{B}^2\rangle,|\mathcal{B}^4\rangle\}$ by exchanging the
outcomes. That is, if their outcomes are anticorrelated, the state
of the whole system is initially in the set
$\{|\mathcal{B}^2\rangle,|\mathcal{B}^4\rangle\}$, otherwise, it is
in the set $\{|\mathcal{B}^1\rangle,|\mathcal{B}^3\rangle\}$. With
this method, Alice and Bob can measure anyone of $n$ pairs of
particles $\{(A_i,B_i),i=1,2,\cdots,n\}$ and identify the states of
the other $n-1$ pairs of particles in the quantum channel of
Eq.~(\ref{Phik2nm}). Then Alice and Bob can realize the
teleportation of an $n$-particle state with a high fidelity when $n$
is large, out of the control of Charlies. Especially, when the
$n$-particle state $|\psi\rangle_{T_1T_2\cdots T_n}$ [see
Eq.~(\ref{psiTn})] is separable, such as $y_{j_1j_2\cdots
j_n}=y_{j_1}y_{j_2}\cdots y_{j_n}$, Alice and Bob can realize
perfect teleportation of $n-1$ qubits information escaping from the
control of Charlies.

However, this drawback can be avoided by the following methods. We
can establish two sequences of states chosen from the four Bell
states for the $n$ pairs of particles $\{(A_i,B_i)\}$, and make
one-to-one correspondence between them and the two states $|00\cdots
0\rangle_{C_1C_2\cdots C_m}$ and
$(1/\sqrt{S_m^{k-1}})|k-1,m-k+1\rangle_{C_1C_2\cdots C_m}$. That is,
we can use the following entangled state, instead of
$|\Phi^k\rangle_{2n+m}$, to act as the quantum channel:
\begin{eqnarray}
\label{Phik2nm1}
 |\Phi'^k\rangle_{2n+m}&=&\frac{1}{\sqrt{2}}\left(\prod\limits_{i=1}^{n}|\mathcal{B}^{r_i}\rangle_{A_iB_i}
      \otimes|00\cdots 0\rangle_{C_1C_2\cdots C_m}\right.\nonumber\\
&&\left.+\prod\limits_{i=1}^{n}|\mathcal{B}^{s_i}\rangle_{A_iB_i}\otimes\frac{1}{\sqrt{S_m^{k-1}}}|k-1,m-k+1\rangle_{C_1C_2\cdots
C_m}\right),
\end{eqnarray}
where $r_i$ ($s_i$) $=1,2,3,$ or $4$. Note that Alice and Bob can
know the $n$ pairs of particles $\{(A_i,B_i)\}$ are in the sequence
of states $\prod\limits_{i=1}^{n}|\mathcal{B}^{r_i}\rangle_{A_iB_i}$
or $\prod\limits_{i=1}^{n}|\mathcal{B}^{s_i}\rangle_{A_iB_i}$ if and
only if they know the particles $\{C_j,j=1,2,\cdots,m\}$ are in the
state $|00\cdots 0\rangle_{C_1C_2\cdots C_m}$ or
$(1/\sqrt{S_m^{k-1}})|k-1,m-k+1\rangle_{C_1C_2\cdots C_m}$ by
Charlies' help. In other words, they cannot ascertain which sequence
of states the subsystem of their $n$ pairs of particles is in
without the cooperation of Charlies. The teleportation of an
arbitrary $n$-particle state can also be implemented by using a
genuine $2n$-particle entangled state as shown in
Refs.~\cite{96PRL060502,74PRA032324,364PLA7}. Thus the quantum
channel of the ($k$, $m$)-threshold CT of an $n$-particle state can
also be constructed as the following form for avoiding the
aforementioned drawback:
\begin{eqnarray}
\label{Phik2nm2}
 |\Phi''^k\rangle_{2n+m}&=&\frac{1}{\sqrt{2}}\left(|MES\rangle_{A_1\cdots A_nB_1\cdots B_n}
      \otimes\frac{1}{\sqrt{S_m^{k-1}}}|k-1,m-k+1\rangle_{C_1C_2\cdots C_m}\right.\nonumber\\
   && \left.+\sigma^{j_1}_{A_1}\cdots\sigma^{j_n}_{A_n}|MES\rangle_{A_1\cdots A_nB_1\cdots B_n}
     \otimes|00\cdots 0\rangle_{C_1C_2\cdots C_m}\right),
\end{eqnarray}
where $|MES\rangle_{A_1\cdots A_nB_1\cdots B_n}$ is a genuine
$2n$-particle entangled state showed in Eq.~(18) of
Ref.~\cite{96PRL060502} (for $n=2$) or Eq.~(10) of
Ref.~\cite{74PRA032324}, $j_i=0,x,y$, or $z$ ($i=1,2,\cdots,n$) with
$\sigma^{0}$ being the two-dimensional identity operator. However,
when $n\geq 4$, $|MES\rangle_{A_1\cdots A_nB_1\cdots B_n}$ was not
explicitly constructed in Ref.~\cite{74PRA032324}. As shown in
Ref.~\cite{364PLA7}, $|MES\rangle_{A_1\cdots A_nB_1\cdots B_n}$ can
be replaced by a $2n$-qubit cluster state
$|Cluster\rangle_{A_1B_1\cdots
A_nB_n}=(|0\rangle_{A_1}+|1\rangle_{A_1}\sigma^z_{B_1})(|0\rangle_{B_1}+|1\rangle_{B_1}\sigma^z_{A_2})\cdots
(|0\rangle_{A_n}+|1\rangle_{A_n}\sigma^z_{B_n})(|0\rangle_{B_n}+|1\rangle_{B_n})$
\cite{86PRL910}. Then the quantum channel reads
\begin{eqnarray}
\label{Phik2nm3}
 |\Phi'''^k\rangle_{2n+m}&=&\frac{1}{\sqrt{2}}\left(|Cluster\rangle_{A_1B_1\cdots A_nB_n}
      \otimes\frac{1}{\sqrt{S_m^{k-1}}}|k-1,m-k+1\rangle_{C_1C_2\cdots C_m}\right.\nonumber\\
   && \left.+\sigma^{j_1}_{A_1}\cdots\sigma^{j_n}_{A_n}|Cluster\rangle_{A_1B_1\cdots A_nB_n}
     \otimes|00\cdots 0\rangle_{C_1C_2\cdots C_m}\right).
\end{eqnarray}
 Note that $\sigma^{j_1}_{A_1}\cdots\sigma^{j_n}_{A_n}$ in
Eqs.~(\ref{Phik2nm2}) and (\ref{Phik2nm3}) can not be set to
$\sigma^{0}_{A_1}\cdots\sigma^{0}_{A_n}$, i.e., $j_i$ can not be
simultaneously equal to zero.

\subsection{The features of the entanglement channels}

It is known that the complexity of multipartite entanglement
increases greatly with the increase of the number of parties
involved. So far, the properties of multipartite entanglement are
not very clear. The classification and quantification of genuine
three-qubit \cite{threequbit} and four-qubit \cite{fourqubit}
entangled states were intensively studied. The classification and
quantification of genuine entangled states involving more than four
qubits were also discussed \cite{multiqubit,74PRA022314}. Although
several typical multipartite entangled states, such as GHZ states
\cite{GHZ}, W states \cite{threequbit}, and cluster states
\cite{86PRL910}, were presented, the inequivalent types of genuine
multipartite entangled states for more than four particles are still
very vague. It will need a long-term effort to well understand the
entanglement involving many parties. To seek for genuine
multipartite entangled states we can resort to particular quantum
schemes since sharing a unique entanglement may allow ones to do
some things that ones cannot otherwise do. Teleportation is a well
example, with which some genuine multipartite entangled states were
found \cite{75PRA052306,96PRL060502,74PRA032324}. Obviously, all the
states $|\Phi^k\rangle_{2+m}$ [see Eq.~(\ref{Phik})],
$|\Phi^k\rangle_{2n+m}$ [see Eq.~(\ref{Phik2nm})],
$|\Phi'^k\rangle_{2n+m}$ [see Eq.~(\ref{Phik2nm1})],
$|\Phi''^k\rangle_{2n+m}$ [see Eq.~(\ref{Phik2nm2})], and
$|\Phi'''^k\rangle_{2n+m}$ [see Eq.~(\ref{Phik2nm3})], which act as
the quantum channels in our ($k$, $m$)-threshold CT schemes, are
genuine multipartite entangled states, because any bipartite cut in
them is inseparable \cite{74PRA022314}. Here, we roughly show the
relationships or differences between them and other genuine
multipartite entangled states presented in literature.

We begin with the state $|\Phi^1\rangle_{2+m}$. When $m=1$,
$|\Phi^1\rangle_{2+1}=(1/2)(|00\rangle_{AB}|+\rangle_C+|11\rangle_{AB}|-\rangle_{C})$
is a three-qubit GHZ state; when $m=2$,
$|\Phi^1\rangle_{2+2}=(1/2)(|0000\rangle+|0011\rangle+|1100\rangle-|1111\rangle)_{ABC_1C_2}$
is just a four-qubit linear cluster state \cite{86PRL910}. As to
$m>2$, $|\Phi^1\rangle_{2+m}=\mathrm{l.u.}|G\rangle_{2+m}$, where
``l.u.'' indicates that the equality holds up to local unitary
transformations on one or more of the qubits and
\begin{eqnarray}
\label{G}
|G\rangle_{2+m}&=&(|0\rangle_A+|1\rangle_A\sigma^z_B)(|0\rangle_B+|1\rangle_B\sigma^z_{C_1})\nonumber\\
 &&\otimes(|0\rangle_{C_1}+|1\rangle_{C_1}\sigma^z_{C_2}\cdots \sigma^z_{C_m})\prod\limits_{i=2}^{m}(|0\rangle_{C_i}+|1\rangle_{C_i})
 \end{eqnarray}
is a graph state \cite{69PRA062311} shown in Fig.~1. Obviously, when
$m>2$, $|\Phi^1\rangle_{2+m}$ is inequivalent to the well-known GHZ,
W, and linear cluster states, in terms of stochastic local
operations and classical communications (SLOCC). By the way, many
schemes for generating multi-qubit graph states were presented (see
e.g., \cite{97PRL143601}), and the six-qubit graph states are
already achievable in the optical experiment \cite{3NP91}.

\begin{figure}[htp]
  \includegraphics[width=9cm,height=3cm]{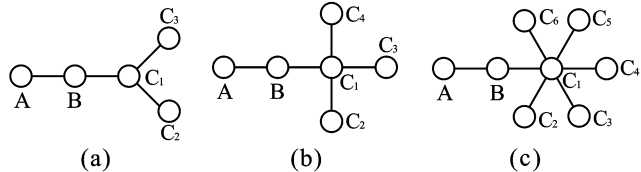}
\caption{The graph state of Eq.~(\ref{G}). (\textbf{a}) $m=3$.
(\textbf{b}) $m=4$. (\textbf{c}) $m=6$.}
\label{figure1}
\end{figure}

 In order to compare the state $|\Phi^k\rangle_{2+m}$ with the corresponding GHZ
state $|GHZ\rangle_{2+m}$, W state $|W\rangle_{2+m}$, and linear
cluster state $|Cluster\rangle_{2+m}$, we resort to the concept of
\emph{persistency of entanglement} \cite{86PRL910}. The
\emph{persistency of entanglement} $P_e(|\Psi\rangle)$  of an
entangled state $|\Psi\rangle$ of $N$ particles is the minimum
number of local measurements such that, for all measurement
outcomes, the state is completely disentangled. For pure states, a
completely disentangled state means a product state of all $N$
particles \cite{86PRL910}. Evidently, for all $N$-qubit states
$0\leq P_e\leq N-1$. As shown in Ref.~\cite{86PRL910}, two states
with different $P_e$ are SLOCC inequivalent, but the inverse case
needs further investigation. We now discuss the three cases as
follows. (a) $k<m$ and $k\neq m/2$. We can prove that
$P_e(|\Phi^k\rangle_{2+m})=k+1$ is different from
$P_e(|GHZ\rangle_{2+m})=1$, $P_e(|W\rangle_{2+m})=m+1$, and
$P_e(|Cluster\rangle_{2+m})=[(m+2)/2]$ \cite{86PRL910}. Thus the
state $|\Phi^k\rangle_{2+m}$ is SLOCC inequivalent to the
corresponding GHZ, W, and linear cluster states. (b) $k= m/2$.
$P_e(|\Phi^k\rangle_{2+m})=m/2+1=P_e(|Cluster\rangle_{2+m})$. The
relation of $|\Phi^k\rangle_{2+m}$ and $|Cluster\rangle_{2+m}$ needs
further investigation. (c) $k=m$.
$P_e(|\Phi^m\rangle_{2+m})=m+1=P_e(|W\rangle_{2+m})$. Then we cannot
distinguish between $|\Phi^m\rangle_{2+m}$ and $|W\rangle_{2+m}$ by
this method. However, we notice that $|\Phi^m\rangle_{2+m}$ belongs
to the GHZ-W-type entangled states recently proposed by Chen
\emph{et al.} \cite{74PRA062310}, and thus does not belong to the
W-type states. On the other hand, $\mathrm{tr}_{C_1\cdots
C_m}(|\Phi^m\rangle_{2+m}\langle\Phi^m|)=\frac{1}{2}|00\rangle_{AB}\langle00|
+\frac{1}{2}|11\rangle_{AB}\langle11|$ is a separable state and
$\mathrm{tr}_{C_1\cdots C_m}(|W\rangle_{2+m}\langle
W|)=\frac{m}{m+2}|00\rangle_{AB}\langle 00|
+\frac{2}{m+2}|\mathcal{B}^3\rangle_{AB}\langle\mathcal{B}^3|$ is a
partially mixed entangled state, which also justifies the conclusion
that $|\Phi^m\rangle_{2+m}$ and $|W\rangle_{2+m}$ are SLOCC
inequivalent. By the way, a scheme for generating a GHZ-W-type state
has been proposed lately \cite{79PRA062315}. Similarly, we can prove
that all the states $\{|\Phi^k\rangle_{2+(m+2n-2)}~[\mathrm{see}~
 \mathrm{Eq.}~(\ref{Phik})],|\Phi^k\rangle_{2n+m}~(\mathrm{or}~
|\Phi'^k\rangle_{2n+m}), |GHZ\rangle_{2n+m}, |W\rangle_{2n+m},
|Cluster\rangle_{2n+m} \}$ are generally SLOCC inequivalent to each
other.

Now, let us pay attention to the states $|\Phi''^k\rangle_{2n+m}$
and $|\Phi'''^k\rangle_{2n+m}$. In the state
$|\Phi''^k\rangle_{2n+m}$, $|MES\rangle_{A_1\cdots A_nB_1\cdots
B_n}$ is explicitly constructed for $n=2$ \cite{96PRL060502} and
$n=3$ \cite{74PRA032324}, respectively. That is,
\begin{eqnarray}
|MES\rangle_{A_1A_2B_1B_2}&=&\frac{1}{2\sqrt{2}}(|0000\rangle-|0011\rangle-|0101\rangle+|0110\rangle\nonumber\\
  && +|1001\rangle+|1010\rangle+|1100\rangle+|1111\rangle)_{A_1A_2B_1B_2},\nonumber\\
|MES\rangle_{A_1A_2A_3B_1B_2B_3}&=&\frac{1}{2\sqrt{2}}
(|000000\rangle +|010110\rangle +|110100\rangle +|100010\rangle\nonumber\\
&&+|011011\rangle+|001101\rangle+|101111\rangle+|111001\rangle)_{A_1A_2A_3B_1B_2B_3}.
\end{eqnarray}
Both the states were proved to be SLOCC inequivalent to the
corresponding GHZ and W states \cite{96PRL060502,74PRA032324}. By
the way, a scheme for generating $|MES\rangle_{A_1A_2B_1B_2}$ has
been proposed recently \cite{78PRA024301}. We notice that
$|\Phi''^1\rangle_{4+1}$ is SLOCC equivalent to the state of
Eq.~(17) of Ref.~\cite{75PRA052306} which was constructed also for
implementing (1,1)-threshold CT of a two-particle state. In
addition, $|\Phi''^1\rangle_{4+2}$ can be transformed into
$|MES\rangle_{A_1A_2A_3B_1B_2B_3}$ by local operations with $C_1$
and $C_2$ replaced by $A_3$ and $B_3$, respectively;
$|\Phi'''^1\rangle_{2n+1}$ is a $(2n+1)$-qubit linear cluster state.
It can be proved that
$P_e(|\Phi''^k\rangle_{2n+m})=P_e(|\Phi'''^k\rangle_{2n+m})=n+k$.
Thus when $k\neq m/2$ ($m>1$), $|\Phi''^k\rangle_{2n+m}$ and
$|\Phi'''^k\rangle_{2n+m}$ are SLOCC inequivalent to the
corresponding GHZ, W, and linear cluster states. As to the case
$k=m/2$, $|\Phi''^k\rangle_{2n+m}$ and $|\Phi'''^k\rangle_{2n+m}$
are also SLOCC inequivalent to the corresponding GHZ and W states,
but the relation of them and linear cluster states needs further
investigation.

\section{Security of the ($k$, $m$)-threshold controlled teleportation}
Our ($k$, $m$)-threshold CT schemes are secure against both Bob's
dishonesty and Charlies' treacheries.

\subsection{Security against Bob's dishonesty}
 Bob may manage to recover
Alice's original state out of the control of Charlies. Thus, during
the distribution of the quantum channel, he intercepts $k$ or more
of the particles $\{C_i,i=1,2,\cdots, m\}$ and performs them
single-particle measurements with the basis
$\{|0\rangle,|1\rangle\}$, and resends them or sends other $k$ or
more auxiliary particles to corresponding Charlies, respectively. By
this way, Bob can ascertain the state of the subsystem of pairs of
particles $\{(A_j,B_j),j=1,2,\cdots\}$ and thus successfully
recovers Alice's original state without the cooperation of Charlies.
However, the correlation among particles $A_j$, $B_j$, and $C_i$ is
disturbed or destroyed. We take $k=1$ as an example. If Bob performs
a measurement on one of the particles $\{C_i\}$ and directly resends
it to corresponding Charlie, the subsystem of Charlies will be in a
product state $|00\cdots 0\rangle_{C_1C_2\cdots C_m}$ or $|11\cdots
1\rangle_{C_1C_2\cdots C_m}$. Then there is no any correlation among
particles $\{C_i\}$. This case can be can be easily found by
Charlies. If Bob sends other $m$ auxiliary particles in a GHZ state
$(1/2)(|00\cdots 0\rangle+|11\cdots 1\rangle)_{C'_1C'_2\cdots C'_m}$
to Charlies, the correlation between the subsystem of Alice and Bob
and that of Charlies is destroyed. Thus such an action of Bob can
also be detected. In fact, the correlation of any genuine
multipartite entangled state will be disturbed or destroyed by any
measurement on a subspace of it, and cannot be perfectly simulated
by another entangled state involving less parties. As a consequence,
Bob's dishonest action can always be detected in our schemes. The
detailed proof is so complicated and prolix, and will be given
elsewhere. Note that Charlies should randomly choose a sufficient
subset of quantum channels to check whether particles are
intercepted during the distribution before carrying out the task of
CT. The security checking process is similar to that of quantum
secret sharing schemes (see, e.g. \cite{41JPA255309}). As a matter
of fact, most of quantum communication schemes need ones to use this
method to check the security of quantum channels against
eavesdropper's interception. Also, all the previous CT schemes
\cite{58PRA4394,70PRA022329,72PRA022338,75PRA052306,0609026,68PRA022321,79PRA062313}
are secure against Bob's dishonesty if checking the security of
quantum channels before carrying out the corresponding tasks.

\subsection{Security against Charlies' treacheries}

When some Charlies are not satisfied with a collective decision,
they may betray the community by three possible ways as follows. (a)
They privately help Bob to reconstruct Alice's original state. (b)
They reject cooperating with Bob and making measurements on their
particles. (c) They cheat Bob and send him the false measurement
outcomes. We assume that any classical communication is open and
insecure, and treacherous Charlies will be punished if their
treacherous actions are detected. Then cases (a) and (b) will not
occur. In the following, we show how case (c) can be prevented.

We first consider that there is only one treacherous Charlie, e.g.,
Charlie $j$, who cheats Bob and sends him the false measurement
outcome. That is, when Charlie $j$ gets the measurement outcome
$|0\rangle$ he broadcasts $|1\rangle$, when getting $|1\rangle$ he
broadcasts $|0\rangle$. There are two cases. Case one: $k<m$. If the
real measurement outcome on the subsystem of particles
$\{C_i,i=1,2,\cdots, m\}$ is $|00\cdots 0\rangle_{C_1C_2\cdots
C_m}$, then the broadcasted outcome is $|00\cdots 010\cdots
0\rangle_{C_1C_2\cdots C_{j-1}C_{j}C_{j+1}\cdots C_m}$ because
Charlie $j$ announced the opposite outcome. However, such an outcome
should not appear when there is no treacherous Charlie. Thus the
cheat action of Charlie $j$ is exposed. If the real measurement
outcome is one term of
$(1/\sqrt{S_m^{k-1}})|k-1,m-k+1\rangle_{C_1C_2\cdots C_m}$ involving
$k-1$ zeros and $m-k+1$ ones, then the broadcasted outcome involves
$k-2$ or $k$ zeros. In this case, Bob can also find that there
exists a betrayer, although he cannot directly know which Charlie
cheated him. In a word, Bob can always detect whether or not there
exist treacherous Charlies who cheat him. The probability of exactly
finding the cheat action of Charlie $j$ is $1/2$. Case two: $k=m$.
If the real measurement outcome is $|00\cdots 0\rangle_{C_1C_2\cdots
C_m}$ or $|00\cdots 010\cdots 0\rangle_{C_1C_2\cdots
C_{j-1}C_{j}C_{j+1}\cdots C_m}$, the broadcasted outcome is
$|00\cdots 010\cdots 0\rangle_{C_1C_2\cdots
C_{j-1}C_{j}C_{j+1}\cdots C_m}$ or $|00\cdots 0\rangle_{C_1C_2\cdots
C_m}$. Then the cheat action of Charlie $j$ cannot be found and Bob
will obtain a wrong state instead of Alice's original state. If the
real measurement outcome is $|10\cdots 0\rangle_{C_1C_2\cdots C_m}$,
then the broadcasted outcome is $|10\cdots 010\cdots
0\rangle_{C_1C_2\cdots C_{j-1}C_{j}C_{j+1}\cdots C_m}$. However,
such an outcome should not appear when there is no treacherous
Charlie. Thus Bob can find that there exists a betrayer. In a
nutshell, the probability of finding the existence of treacherous
Charlie is $(S_m^{k-1}-1)/(2S_m^{k-1})$.  Note that they may
randomly broadcast an artificial outcome without measurement. This
way has no essential differences with the one discussed above.

For the case where there are $l$ ($l<m$) treacherous Charlies who
send the false outcomes to Bob, when $l\neq m-k+1$, the probability
of finding the treacherous Charlies is one ($l$ is odd) or
$1-S_l^{l/2}/S_m^k$ ($l$ is even); when $l=m-k+1$, the probability
is $(S_m^k-1)/(2S_m^k)$ ($l$ is odd) or
$(S_m^k-S_l^{l/2}-1)/(2S_m^k)$ ($l$ is even).

According to the above analysis, when there is only one Charlie who
cheats Bob and sends him the false measurement outcome, his cheat
action can be directly detected with probability $1/2$. Because when
the cheat action of any one of Charlies is found, he will be
chastised, the case where one or more Charlies cheat Bob will not
occur in practice. We now prove it by the game theory \cite{game}.
Assume that there are $l$ potential treacherous Charlies who are not
satisfied with a collective decision that permitting Bob to
reconstruct Alice's original state. They will play a multi-player
\emph{Prisoners-Dilemma-like} game. The so-called Prisoners' Dilemma
game \cite{prisoner} is as follows. Two or more perpetrators are
caught by the police and are interrogated in separate cells
\emph{without communication among them}. Unfortunately, the police
lacks enough proof to implead them. The chief policeman now makes
the following offer to each prisoner: if one of them confesses to
the crime, but the others do not, then he or she will be commuted by
$r$ years and the others will increase $r$ years; if all of them
deny, then each of them will be commuted by $s$ years ($s<r$); if
all of them confess, then everyone will be commuted by $t$ years
($t<s<r$). The objective of each player (prisoner) is to maximize
his or her individual payoff. The catch of the dilemma is that
confessing (i.e., they defect from each other) is the dominant
strategy, that is, rational reasoning forces each player to defect,
and thereby doing substantially worse than if they would all decide
to cooperate (deny). In terms of the game theory, such a mutual
defection is a Nash equilibrium \cite{Nash} because each of the
players comes to the conclusion that he or she could not have done
better by unilaterally changing his or her own strategy. In our
scheme, if one of the potential treacherous Charlies sends Bob the
false outcome, and the others do not, he will be detected and
chastised and they will achieve their purpose of preventing Bob from
recovering Alice's original state; if two or more of them send false
outcomes, they can accomplish their purpose escaping from penalty;
if all of them do not send false outcome, each will not be punished
but they cannot achieve their aim. Thus each of potential
treacherous Charlies wish their partners but not himself to send the
false outcomes, because then he can accomplish his purpose but not
be chastised. The rational reasoning and selfish gene force each
Charlie to send correct outcome. This decision is a Nash equilibrium
because each of Charlies could not do better by unilaterally
changing his action.

In a word, our schemes are secure against Charlies' cheats. It is
worth pointing out that all previous CT schemes
\cite{58PRA4394,70PRA022329,72PRA022338,75PRA052306,40JPB1767,0609026},
including the scheme of Ref.~\cite{79PRA062313}, are insecure when
there exist treacherous Charlies. That is, the cheat action of
Charlies can not be detected. Then Bob may obtain a wrong state with
very low fidelity instead of Alice's original state when one or more
Charlies send him the false measurement outcomes. For instance, we
consider the CT of a single-particle state $|\psi\rangle_T$ [see
Eq.~(\ref{psi})] with a standard GHZ state. When there are odd
Charlies who send the false measurement outcomes to Bob, he will get
a wrong state with only the fidelity $F=(|\alpha|^2-|\beta|^2)^2$.

\section{Concluding remarks}

In summary, we have proposed several ($k$, $m$)-threshold
controlling schemes for CT, where the teleportation of a quantum
state Alice to Bob is under the control of $m$ Charlies such that
$k$ ($k\leq m$) or more of them can help Bob successfully recover
the transferred state. We have also shown that our schemes are
secure against both Bob's dishonesty and Charlies' treacheries.
However, previous ($m$, $m$)-threshold schemes cannot prevent
Charlies' cheats. The presented schemes have potential applications
in networked quantum information processing. For example, they can
be used to implement the ($k$, $m$)-threshold quantum-secret-sharing
without nonlocal operation among receivers and additional limitation
for $k$, following the idea of Ref.~\cite{79PRA062313}. Our schemes
are also useful to seek and explore genuine multipartite entangled
states. We utilized the game theory to prove the security of our
schemes against Charlies' cheats. This implies that our schemes may
open another perspective for the applications of the game theory.

Although we only discussed the case where the quantum channels are
pure entangled states, suitable mixed entangled states may also be
competent for the ($k$, $m$)-threshold CT. In fact, the general form
of the pure-entangled-state channel of Eq.~(\ref{Phi}) can be
replaced by the mixed-state channel
\begin{eqnarray}
\label{rho}
 \rho_{2+m}&=&|x_1|^2|\mathcal{B}^1\rangle_{AB}\langle\mathcal{B}^1|\otimes|\phi^1\rangle_{C_1C_2\cdots
        C_m}\langle\phi^1|\nonumber\\
     && + |x_2|^2|\mathcal{B}^2\rangle_{AB}\langle\mathcal{B}^2|\otimes|\phi^2\rangle_{C_1C_2\cdots
        C_m}\langle\phi^2|\nonumber\\
  && +|x_3|^2|\mathcal{B}^3\rangle_{AB}\langle\mathcal{B}^3|\otimes|\phi^3\rangle_{C_1C_2\cdots
        C_m}\langle\phi^3|\nonumber\\
  && +|x_4|^2|\mathcal{B}^4\rangle_{AB}\langle\mathcal{B}^4|\otimes|\phi^4\rangle_{C_1C_2\cdots
        C_m}\langle\phi^4|.
\end{eqnarray}
Then corresponding mixed-state channels of the ($k$, $m$)-threshold
CT can be constructed by the same methods as in Sec. II B and Sec.
II C. With the forms of the states of Eqs.~(\ref{Phi}) and
(\ref{rho}), one can construct different quantum channels for
implementing ($k$, $m$)-threshold CT. Note that all the quantum
channels should at least satisfy the following conditions. (a) They
are symmetric under permutation of qubits $\{C_1,C_2,\cdots, C_m\}$.
(b) The four states $\{|\phi^1\rangle_{C_1C_2\cdots
C_m},|\phi^2\rangle_{C_1C_2\cdots C_m},|\phi^3\rangle_{C_1C_2\cdots
C_m},|\phi^4\rangle_{C_1C_2\cdots C_m}\}$ can not be fully
distinguished unless $k$ of supervisors perform single-particle
measurements on their own particles with appropriate bases and
combine the measurement outcomes. In addition, different methods may
be needed to discuss the security of concrete schemes.

As mentioned above, SaiToh \emph{et al.} \cite{79PRA062313} also
proposed a ``($k$, $m$)-threshold'' CT scheme which is a combination
of a ($m$, $m$)-threshold CT scheme and a ($k$, $m$)-threshold
secret sharing scheme. In their scheme, however, the receiver Bob
still needs receiving all of the supervisors' correct measurement
outcomes, i.e., needs the cooperation of all Charlies, for
recovering the teleported state. Thus their scheme is not a genuine
($k$, $m$)-threshold controlling scheme and can not prevent
Charlies' cheats. They also mentioned that a ($k$, $m$)-threshold
controlling scheme can be constructed by sharing a classical key
among Charlies such that $k$ or more of them can recover the key.
The distribution of the key can be achieved by quantum cryptography.
However, they did not construct a concrete scheme. In addition, as
shown in Ref.~\cite{79PRA062313}, a classical key can be easily
copied, and Charlies cannot stop Bob from recovering Alice's
original state if Bob manages to obtain as least $k$ shares of the
key without consent of Charlies. More importantly, the classical
($k$, $m$)-threshold controlling scheme can not prevent Charlies'
cheats. In principle, a ($k$, $m$)-threshold controlling scheme can
be constructed by using the quantum polynomial codes \cite{83PRL648}
as mentioned in Ref.~\cite{79PRA062313}. However, it needs Charlies
and Bob to come together and perform nonlocal operations
(multi-particle operations). In contrast, our schemes do not need
performing nonlocal operations and are secure against Charlies'
cheats of
sending false measurement outcomes.\\

\section*{Acknowledgements}
This work is supported by National Natural Science Foundation of
China, Project No. 10674018 and No. 10874019, and the National
Fundamental Research Program of China, Projects No. 2004CB719903.


\begin{thebibliography}{}
\bibitem{70PRL1895} Bennett C H, Brassard G, Cr\'{e}peau C, Jozsa R, Peres A and Wootters W K 1993 \emph{Phys.
Rev. Lett.} \textbf{70} 1895
\bibitem{primitive} Gottesman D and Chuang I L 1999 \emph{Nature} \textbf{402} 390\\
 Knill E, Laflamme L and Miburn G J 2001 \emph{Nature} \textbf{409} 46\\
 Kok P, Munro W J, Nemoto K, Ralph T C, Dowling J P and Milburn G J 2007 \emph{Rev. Mod. Phys.} \textbf{79} 135
\bibitem{390N575} Bouwmeester D, Pan J W, Mattle K, Eibl M, Weinfurter H and Zeilinger A 1997 \emph{Nature} \textbf{390} 575\\
Marcikic I, de Riedmatten H, Tittel W, Zbinden H and Gisin N 2003 \emph{Nature} \textbf{421} 509\\
 Chen Y A, Chen S, Yuan Z S, Zhao B, Chuu C S, Schmiedmayer J and Pan J W 2008 \emph{Nat. Phys.} \textbf{4} 103\\
 Olmschenk S, Matsukevich D N, Maunz P, Hayes D, Duan L M and Monroe C 2009 \emph{Science} \textbf{323} 486
\bibitem{58PRA4394} Karlsson A and Bourennane M 1998 \emph{Phys. Rev.} A \textbf{58} 4394
\bibitem{70PRA022329} Yang C P, Chu S and Han S 2004 \emph{Phys. Rev.} A \textbf{70} 022329\\
 Yang C P, Han S 2005 \emph{Phys. Lett.} A \textbf{343} 267
\bibitem{72PRA022338} Deng F G, Li C Y, Li Y S, Zhou H Y and Wang Y 2005 \emph{Phys. Rev.} A \textbf{72} 022338
\bibitem{40JPB1767} Man Z X, Xia Y J and An N B 2007 \emph{J. Phys. B: At. Mol. Opt. Phys.} \textbf{40} 1767\\
 Li X H, Zhou P, Li C Y, Zhou H Y and Deng F G 2006 \emph{J. Phys. B: At. Mol. Opt. Phys.} \textbf{39} 1975
\bibitem{75PRA052306} Man Z X, Xia Y J and An N B 2007 \emph{Phys. Rev.} A \textbf{75} 052306
\bibitem{0609026} Kenigsberg D and Mor T 2006 \emph{Preprint} quant-ph/0609028
\bibitem{68PRA022321} An N B 2003 \emph{Phys. Rev.} A \textbf{68} 022321
\bibitem{341PLA55} Zhang Z J and Man Z X 2005 \emph{Phys. Lett.} A \textbf{341} 55\\
   Zhang Z J 2006 \emph{Phys. Lett.} A \textbf{352} 55
\bibitem{430N54} Zhao Z, Chen Y A, Zhang A N, Yang T, Briegel H J and Pan J W 2004 \emph{Nature} \textbf{430} 54
\bibitem{59PRA1829} Hillery M, Bu\v{z}ek V and Berthiaume A 1999 \emph{Phsy. Rev.} A \textbf{59} 1829\\
   Deng F G, Li X H, Li C Y, Zhou P and Zhou H Y 2005 \emph{Phys. Rev.} A \textbf{72} 044301\\
  Markham  D and Sanders B C 2008 \emph{Phys. Rev.} A \textbf{78} 042309
\bibitem{79PRA062315} Wang X W and Yang G J 2009 \emph{Phys. Rev.} A \textbf{79} 062315
\bibitem{79PRA062313} SaiToh A, Rahimi R and Nakahara M 2009 \emph{Phys. Rev.} A \textbf{79} 062313
\bibitem{08011544} Ben-Or M, Cr\'{e}peau C, Gottesman D, Hassidim A and Smith A 2006 \emph{Proc. 47th Annual IEEE Symposium on the Foundations
of Computer Science} (FOCS '06) p249-260 (IEEE Press)
\bibitem{0401076} Aoun B and Tarifi M 2004 \emph{Preprint} quant-ph/0401076
\bibitem{cryptography} Biham E, Huttner B and Mor T 1996 \emph{Phys. Rev.} A \textbf{54} 2651\\
   Townsend P D 1997 \emph{Nature} \textbf{385} 47\\
  Bose S, Vedral V and Knight P L 1998 \emph{Phys. Rev.} A \textbf{57} 822
\bibitem{Wang} Wang X W and Yang G J 2009 \emph{Quantum Inf. Process.} \textbf{8} 319
\bibitem{83PRL648} Cleve R, Gottesman D and Lo H K 1999 \emph{Phys. Rev. Lett.} \textbf{83} 648
\bibitem{60PRA1888} Horodecki M, Horodecki P and Horodecki R 1999 \emph{Phys. Rev.} A \textbf{60} 1888\\
 Lee J and Kim M S 2000 \emph{Phys. Rev. Lett.} \textbf{84} 4236\\
 Bandyopadhyay S and Sanders B C 2006 \emph{Phys. Rev.} A \textbf{74} 032310
\bibitem{GHZ} Greenberger D M, Horne M A, Shimony A and Zeilinger A 1990 \emph{Am. J. Phys.} \textbf{58} 1131
\bibitem{103PRL020503} Prevedel R, Cronenberg G, Tame M S, Paternostro M, Walther P, Kim M S and Zeilinger A 2009 \emph{Phys. Rev. Lett.} \textbf{103} 020503\\
 Wieczorek W, Krischek R, Kiesel N, Michelberger P, T\'{o}th G and Weinfurter H 2009 \emph{Phys. Rev. Lett.} \textbf{103} 020504
\bibitem{0609129} Zhang Q, Goebel A, Wagenknecht C, Chen Y A, Zhao B, Yang T, Mair A, Schmiedmayer J and Pan J W 2006 \emph{Nat. Phys.} \textbf{2} 678
\bibitem{96PRL060502} Yeo Y and Chua W K 2006 \emph{Phys. Rev. Lett.} \textbf{96} 060502
\bibitem{74PRA032324} Chen P X, Zhu S Y and Guo G C 2006 \emph{Phys. Rev.} A \textbf{74} 032324
\bibitem{364PLA7} Wang X W, Shan Y G, Xia L X, Lu M W 2007 \emph{Phys. Lett.} A \textbf{364} 7
\bibitem{86PRL910} Briegel H J and Raussendorf R 2001 \emph{Phy. Rev. Lett.} \textbf{86} 910
\bibitem{threequbit} D\"{u}r W, Vidal G and Cirac J I 2000 \emph{Phys. Rev.} A \textbf{62} 062314\\
   Cornelio M F and de Toledo Piza A F R 2006 \emph{Phys. Rev.} A \textbf{73} 032314 \\
 Ac\'{\i}n A, Andrianov A, Costa L, Jan\'{e} E, Latorre J I and Tarrach R 2000 \emph{Phys. Rev. Lett.} \textbf{85} 1560
\bibitem{fourqubit} Verstraete F, Dehaene J, De Moor B and Verschelde H 2002 \emph{Phys. Rev.} A \textbf{65} 052112\\
 Lamata L, Le\'{o}n J, Salgado D and Solano E 2007 \emph{Phys. Rev.} A \textbf{75} 022318\\
 Li D, Li X, Huang H and Li X 2007 \emph{Phys. Rev.} A \textbf{76} 052311
\bibitem{multiqubit} Osterloh A and Siewert J 2005 \emph{Phys. Rev.} A \textbf{72} 012337\\
 Lamata L, Le\'{o}n J, Salgado D and Solano E 2006 \emph{Phys. Rev.} A \textbf{74} 052336
\bibitem{74PRA022314} Rigolin G, de Oliveira T R and de Oliveira M C 2006 \emph{Phys. Rev.} A \textbf{74} 022314
\bibitem{69PRA062311} Hein M, Eisert J and Briegel H J 2004 \emph{Phys. Rev.} A \textbf{69} 062311\\
 Schlingemann D and Werner R F 2001 \emph{Phys. Rev.} A  \textbf{65} 012308
\bibitem{97PRL143601} Bodiya T P and Duan L M 2006 \emph{Phys. Rev. Lett.} \textbf{97} 143601\\
    Browne D E and Rudolph T 2005 \emph{Phys. Rev. Lett.} \textbf{95} 010501\\
 Nielsen M A 2004 \emph{Phys. Rev. Lett.} \textbf{93} 040503
\bibitem{3NP91} Lu C Y, Zhou X Q, G\"{u}hne O, Gao W B, Zhang J, Yuan Z S, Goebel A, Yang T and Pan J W 2007 \emph{Nat. Phys.} \textbf{3} 91
\bibitem{74PRA062310} Chen L and Chen Y X 2006 \emph{Phys. Rev.} A \textbf{74} 062310
\bibitem{78PRA024301} Wang X W and Yang G J 2008 \emph{Phys. Rev.} A \textbf{78} 024301
\bibitem{41JPA255309} Chi D P, Choi J W, Kim J S, Kim T and Lee S 2008 \emph{J. Phys. A: Math. Theor.} \textbf{41} 255309
\bibitem{game} von Neumann J and Morgenstern O 1947 \emph{ The Theory of
Games and Economic Behaviour} (Princeton University Press,
Princeton)
\bibitem{prisoner} Dawkins R 1976 \emph{The Selfish Gene} (Oxford University Press, Oxford)
\bibitem{Nash} Myerson R B 1991 \emph{Game Theory: An Analysis of Conflict} (MIT Press, Cambridge)\\
 Nash J 1950 \emph{Proc. Nat. Acad. Sci.} \textbf{36} 48
\end{thebibliography}
\end{document}